# Power consumption prediction for steel industry

**WT Al-shaibani.Tareq Babaqi and Abdulraqeeb Alsarori (al-shaibani18@itu.edu.tr)**

## Abstract

The use of steel is essential in many industries, including infrastructure, transportation, and modern architecture. Predicting power consumption in the steel industry is crucial to meet the rising demand for steel and promoting city development. However, predicting energy consumption in the steel industry is challenging due to several factors, such as the type of steel produced, the manufacturing process, and the efficiency of the manufacturing facility. This research aims to contribute by creating a predictive model that estimates power consumption in the steel industry. The unique approach combines linear regression to predict a continuous variable related to power consumption and the KNN clustering method to identify the demanding load type. This study's novelty lies in the development of a model that accurately predicts energy consumption in the steel industry, leading to more sustainable and efficient practices. This research contributes to enabling industries to anticipate and optimize their energy consumption, leading to more sustainable practices and economic development.

## Keywords

## 1. Introduction

For many years, accurate prediction of energy consumption in the industrial sector has been a critical issue. Precise predictions can help industries manage their energy usage, reduce costs, and increase efficiency. Historically, statistical models were used to forecast industrial energy consumption by identifying trends and patterns in historical data. However, these models were often inaccurate as they did not account for changes in production levels or weather conditions. In recent years, machine learning algorithms have become popular for energy consumption prediction due to their ability to analyze large datasets and make more precise predictions. These algorithms can be trained on various factors, such as production levels, weather conditions, and equipment efficiency, to forecast energy consumption accurately. In addition, physics-based models that simulate industrial energy consumption by taking into account principles of heat transfer, thermodynamics, and fluid dynamics have also been used. Still, they require significant computational resources and expertise to develop and implement. Overall, industrial energy consumption prediction has made significant progress, and advanced technologies such as machine learning are improving the accuracy and reliability of energy consumption predictions.

This research paper aims to create a predictive model for estimating power consumption in the steel industry. The research team used linear regression and found that $CO_2$ was the dominant feature, with lagging current coming in second, to predict power consumption usage successfully. The research is structured into several sections, including an introduction that provides a brief overview and in-depth insight into the research problem. This is followed by sections on the data and models used to accomplish the work, with Section 4 discussing the results. An evaluation of the findings is presented before concluding the research.

## 2. Data Description

This section discusses the process of gathering and processing datasets used for the machine learning model. The data was obtained by querying the cloud-based information storage system owned and operated by Korea Electric Power Corporation [2]. The dataset comprises daily, monthly, and annual data and was used to compile the final dataset. Standard practice in machine learning involves dividing the data into two sets: a training set and a test set. The machine learning model is trained on the training set, while the performance of the model after training is evaluated using the test set. Table 1 provides an overview of the dataset features.

The first step in data processing was to check for null values. The null values were counted and visualized to ensure they were accurately identified. Furthermore, a final check was performed to verify that question marks were correctly identified as null values. Figure 2.1 shows that the dataset features contain no missing values.

A normality check was performed to determine whether the dataset was normally distributed. A histogram was plotted to assess the distribution of the data visually. The normal distribution is essential because many statistical tests and methods assume normally distributed data. If the data is not normally distributed, the results of these tests and methods may be inaccurate. Figure 2.2 depicts the histogram plots for the dataset features.

Table 2.1: Dataset feature description

| Feature | Description |
| --- | --- |
| Date | Data collected in real time on the first of the month |
| Usage_kWh | Energy Consumption in Industry kWh continuous |
| Lagging Current | Reactive energy kVarh Continuous |
| Leading Current | Reactive energy kVarh Continuous |
| CO2 | CO2 Continuous ppm |
| NSM | Minutes and seconds since midnight S Continuous |
| Week status | Weekday or Weekend |
| Day of week | Sunday, Monday ..etc |
| Load Type | Light Load, Medium Load, Maximum Load |

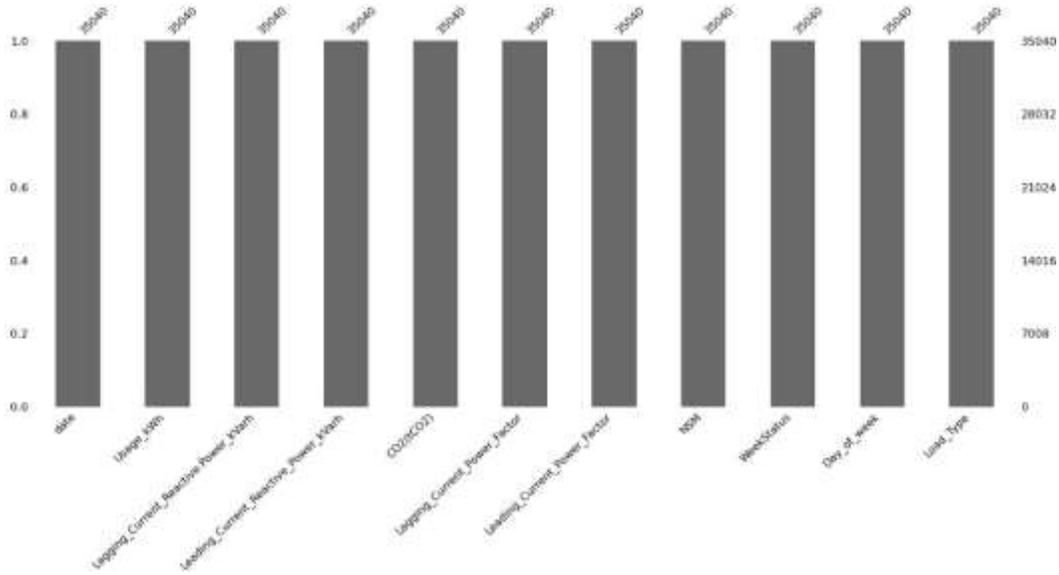

Figure 2.1: The dataset features contains no missing values.

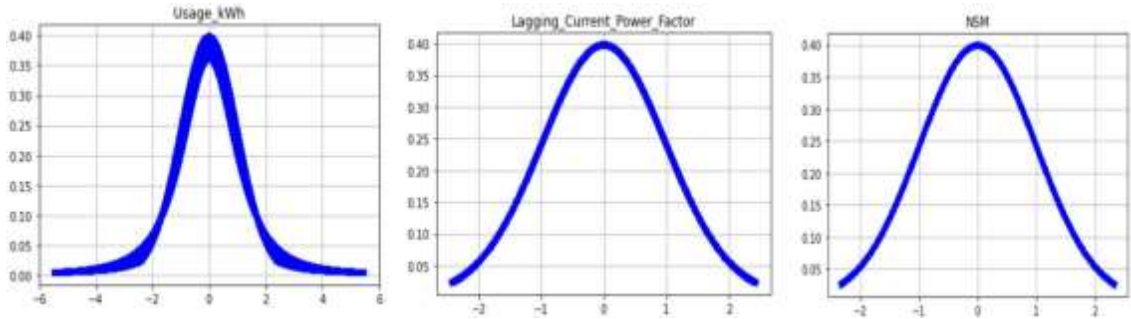

Figure 2.2: Normality check for dataset features

A heatmap was used to visualize the relationship between different variables in the dataset. Heatmapping is a graphical representation of data where the individual values in a matrix are represented as colors. The correlation heatmap displays the magnitude of a linear relationship that exists between two or more variables. Figure 2.3 shows the correlation between different variables with respect to power usage. The highest correlation was found between CO2 and lagging current features because of their relation to power usage.

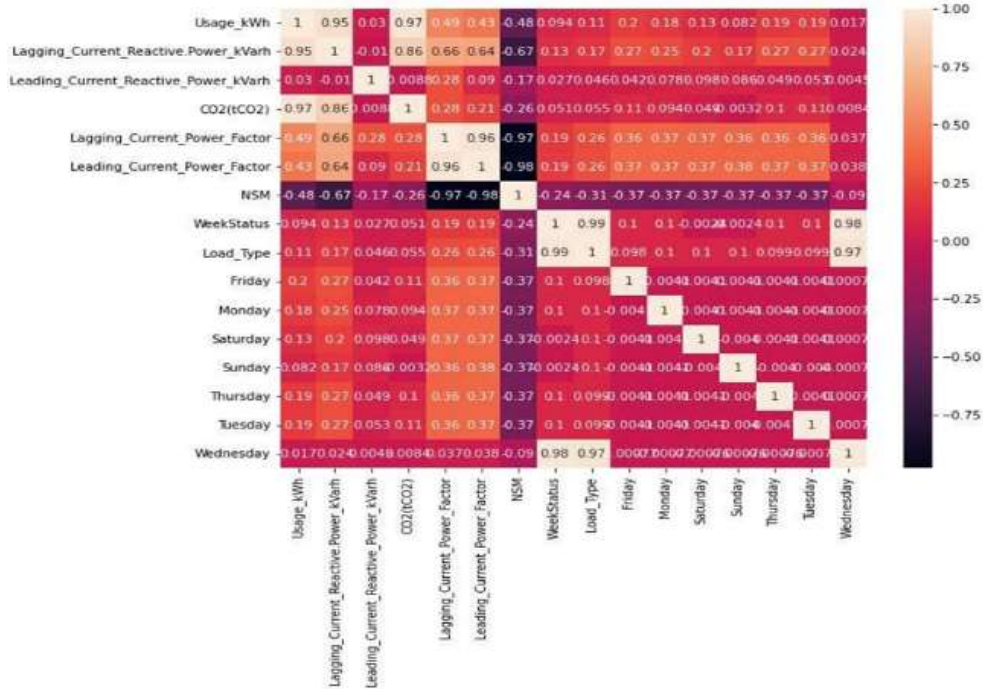
Figure 2.3: The correlation relation between each two features.

## 3. Models

Since the objective of this research is to predict a continuous variable, namely energy consumption, and to identify the type of load, the continuous variable is energy consumption. The best model for this type of problem is typically linear regression, and the KNN model has also been applied to determine the load type [3].

### 3.1. Linear regression

Linear regression is a statistical method used to model the relationship between a dependent variable and one or more independent variables. It is a type of regression analysis that utilizes a straight line to estimate the relationship between a dependent variable and one or more independent variables. On the basis of the values of the independent variables, the equation of the line is used to make predictions about the dependent variable. The dependent variable in linear regression is continuous, meaning it can take any value within a range. The independent variables can be of any type, including continuous or categorical variables.

$$y \equiv \beta_0 + \beta_1 X_i \qquad (3.1)$$

Where y is the dependent variable, $X_i$ is the independent variables. B0 is the intercept, and B1 is the coefficients or slops.

### 3.2. Linear regression with normalization

The process of rescaling data so that it has a mean of 0 and a standard deviation of 1 is known as normalization. If the variables in the dataset have different scales, this can help to make the regression results more interpretable. To perform linear regression with normalization, the data must first be normalized using the following formula 3.2:

$$x\_\text{norm} = \frac{x - \text{mean}(x)}{\text{std}(x)} \qquad (3.2)$$

where "x" is the original variable, "mean(x)" is the variable's mean, and "std(x)" is the variable's standard deviation. Once the data has been normalized, we can perform linear regression as usual, with the normalized variables serving as the independent and dependent variables. Line equation 3.1 will be the same as in regular linear regression, but it will be a but normalized data rather than original data.

### 3.3. Ridge regression

Ridge regression is a regression analysis that is used to examine the relationship between one or more independent variables and a dependent variable. It is a regularized linear regression in which a penalty term is introduced into the model to prevent overfitting and improve the findings' interpretability. The purpose of ridge regression is to discover the coefficient values, or "weights" assigned to each independent variable, that minimize the residual sum of squares, or the sum of the squared differences between the observed and predicted values, subject to a limit on the size of the coefficients. This limitation is known as the "shrinkage penalty," and it is represented by the word "λ" in the ridge regression equation. Ridge regression coefficients are determined by solving the optimization problem shown in 3.3:

$$min[\Sigma(y - x\beta)^2 + \lambda\Sigma\beta^2] \qquad (3.3)$$

Ridge regression is beneficial because, by putting a penalty on the magnitude of the coefficients, it can help to prevent overfitting and enhance model interpretability. This can make the results more stable and dependable and make the link between the variables in the model clearer to grasp.

### 3.4. LASSO

The Least Absolute Shrinkage and Selection Operator is a regularized regression analysis technique used to investigate the relationship between a dependent variable and one or more independent variables. It is similar to ridge regression in that the penalty term is the total of the absolute values of the coefficients rather than the sum of the squared coefficients. The LASSO regression coefficients are determined by solving the optimization problem shown in 3.4:

$$min[\Sigma(y - X\beta)^2 + \lambda\Sigma|\beta|] \qquad (3.4)$$

LASSO regression is effective in determining which independent variables are most essential in predicting the dependent variable. In the optimization problem, the penalty term promotes the coefficients of less relevant variables to be decreased to zero, essentially eliminating them from the model. This can help to make the model more interpretable and prevent overfitting.

### 3.5. KNN

KNN is an abbreviation for "K-Nearest Neighbors." It is an instance-based or lazy learning method in which the function is only approximated locally and all computation is postponed until classification. In other words, the algorithm does not explicitly develop a model, but saves the training data and waits for a prediction request. When a prediction is required, the algorithm locates the nearest training data points (the "nearest neighbors") and makes the forecast using their labels. It is a straightforward and efficient approach to classification and regression problems.

## 4. Results and Discussion

This section shows the outcomes from training the models stated in the Models section. Each model has been trained and its outcomes have been recorded, thus this part will exhibit and discuss these results.

The CO2 was shown to be the most important feature in the linear regression. The correlation value from the heatmap in figure 2.3 might be used to double-check this conclusion. The feature coefficient values are shown in Table 4.1

Table 4.1: Linear regression coefficients

| Feature | Coefficient value |
| --- | --- |
| Lagging Current Reactive Power kVarh | 0.289401 |
| Leading Current Reactive Power kVarh | 0.116304 |
| **CO2(tCO2)** | **1694.237843** |
| Lagging Current Power Factor | 0.114722 |
| Leading Current Power Factor | 0.064895 |
| NSM | 0.000004 |
| Week Status | 0.089524 |
| Load Type | 0.458500 |
| Friday | -0.130716 |
| Monday | -0.118452 |
| Saturday | -0.102274 |
| Sunday | 0.012750 |
| Thursday | -0.143821 |
| Tuesday | 0.528630 |
| Wednesday | -0.046117 |

CO2 was also shown to be an important feature in the linear regression with normalization, but with less importance compared to the previous result. The feature coefficient values are shown in Table 4.2. As previously mentioned, ridge is a shrinking method, and Figure 4.1 depicts the shrinkage process as the penalty increases and the coefficient decreases towards zero. LASSO, on the other hand, displays more interesting results. It shows that CO2 is the most significant contributor, followed by lagging current reactive power. Table 4.4 displays the feature coefficient values for LASSO. Figure 4.2 shows the shrinkage process as the penalty increases and the coefficient decreases towards zero, as previously mentioned.

Table 4.2: Linear regression with normalization coefficients

| Feature | Coefficient value |
|---|---|
| Lagging Current Reactive Power kVarh | 0.500906 |
| Leading Current Reactive Power kVarh | -0.455309 |
| **CO2(tCO2)** | **1413.114801** |
| Lagging Current Power Factor | 0.064763 |
| Leading Current Power Factor | -0.035739 |
| NSM | -0.007579 |
| Week Status | -1.212324 |
| Load Type | 1.324652 |
| Friday | -0.164516 |
| Monday | -0.285530 |
| Saturday | -1.349013 |
| Sunday | -1.467378 |
| Thursday | -0.179278 |
| Tuesday | -0.457353 |
| Wednesday | -0.125647 |

Table 4.3: Ridge regression coefficients

| Feature | Coefficient value |
|---|---|
| **Lagging Current Reactive Power kVarh** | **1.630267** |
| Leading Current Reactive Power kVarh | -0.813496 |
| CO2(tCO2) | 0.015020 |
| Lagging Current Power Factor | 0.140572 |
| Leading Current Power Factor | -0.141011 |
| NSM | -0.013395 |
| Week Status | -0.111796 |
| Load Type | 0.013755 |
| Friday | -0.011596 |
| Monday | -0.001977 |
| Saturday | 0.014301 |
| Sunday | 0.099641 |
| Thursday | -0.074901 |
| Tuesday | -0.096383 |
| Wednesday | 0.073061 |

Table 4.4: LASSO coefficients

| Feature | Coefficient value |
|---|---|
| **Lagging Current Reactive Power kVarh** | **0.003053** |
| Leading Current Reactive Power kVarh | 0.000000 |
| **CO2(tCO2)** | **0.004998** |
| Lagging Current Power Factor | 0.000000 |
| Leading Current Power Factor | 0.000000 |
| NSM | 0.000000 |
| Week Status | 0.000000 |
| Load Type | 0.000000 |
| Friday | 0.000000 |
| Monday | 0.000000 |
| Saturday | 0.000000 |
| Sunday | 0.000000 |
| Thursday | 0.000000 |
| Tuesday | 0.000000 |
| Wednesday | 0.000000 |

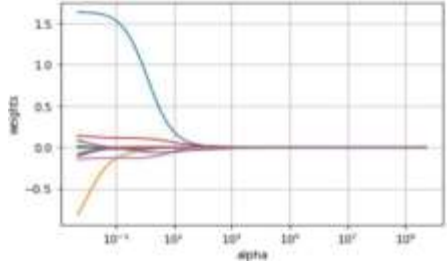
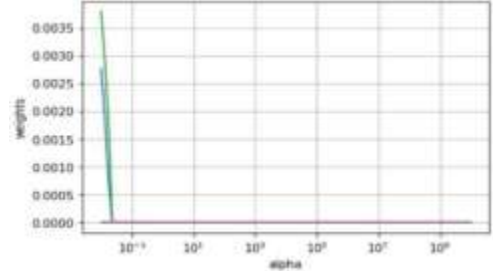

Figure 4.1: Ridge shrinkage process.  Figure 4.2: LASSO shrinkage process.

The KNN methodology has been applied in this instance. First, the optimal value for k was determined by iterating the problem 40 times and selecting the solution with the least error. The load type had then been predicted using the KNN approach, with the optimal founded value of k being utilized in the process. Figure 4.3: illustrates the KNN process to find the optimal value of k [4-6].

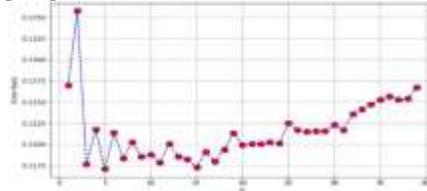

Figure 4.3: Illustrates the Error Rate vs. K Value

## 5. Evaluation

In the field of machine learning, regression is a task that involves predicting a continuous value based on input data, and KNN and linear regression are algorithms that are frequently used for this purpose. To assess the efficacy of a linear regression model's predictions, several metrics such as mean absolute error, mean squared error, and R-squared can be utilized. The mean absolute error (MAE) measures the average magnitude of errors in a model's predictions, without considering the direction of the errors. It is computed as the average of the absolute discrepancies between the anticipated and actual values.

The mean squared error (MSE) is similar to the MAE, but it takes into account the direction of errors. It is determined by taking the average of the squares of the discrepancies between the anticipated and actual values. The R-squared statistic, also known as the coefficient of determination, evaluates a model's ability to predict future outcomes. It represents the fraction of the total variance in the true values that can be accounted for by the model. An R-squared value of 0 indicates that the model does not explain any of the variation, while a value of 1 indicates that the model correctly explains all of the variance in the true values.

These measures can be used to compare the performance of different linear regression models or to track the performance of a single model over time as it is trained and improved. Table 1 presents the evaluation of the different models used in this project. It is clear that the Ridge model has the best performance among all of the models used.

Table 4.4: Evaluation metrics for the models used in the project

|       | Accuracy Score | Mean Absolute Error | Mean Squared Error | Root Mean Squared Error |
|-------|----------------|---------------------|--------------------|-------------------------|
| LR    | 0.981299       | 2.529006            | 20.516540          | 4.529519                |
| **Ridge** | **0.992049** | **0.000143**     | **0.000001**       | **0.001032**            |
| LASSO | 0.978588       | 0.000367            | 0.000003           | 0.001693                |
| KNN   | 0.917333       | 0.089612            | 0.103501           | 0.321715                |

## 6. Conclusions

In conclusion, our study demonstrates that linear regression is an effective approach for estimating power consumption, and the KNN model has been used to accurately categorize different types of loads. When evaluating the models used in this project, it is clear that Ridge regression had the best performance based on the chosen evaluation measures. Our results also indicate that $CO_2$ is the most important feature for predicting power consumption, followed by lagging current. This study has important implications for the energy sector, as accurate power consumption prediction can help improve energy management and reduce costs. Our findings can also serve as a basis for further research and development of more accurate and efficient energy forecasting models.